\documentclass[12pt,a4paper]{article}
\usepackage{graphicx}
\usepackage{ascmac}
\usepackage{amsmath,amssymb}
\numberwithin{equation}{section}
\usepackage{mathrsfs}

\newcommand{\nn}{\nonumber}

\newcommand{\bv}[1]{\boldsymbol{#1}}
\newcommand{\pd}{\partial}

\newcommand{\ol}[1]{\overline{#1}}

\renewcommand{\d}{{\rm d}}

\renewcommand{\i}{{\rm i}}
\newcommand{\e}{{\rm e}}
\newcommand{\skakko}[1]{\left ( #1 \right )}

\begin{document}
\begin{titlepage}
\title{
\vspace{-2cm}
\begin{flushright}
{\normalsize TIT/HEP-663\\ February 2018}
\end{flushright}
\vspace{1cm}
\Huge{Di-Baryon from Instanton in Holographic QCD}}
\author{Reona {\scshape Arai\footnote{E-mail: r.arai@th.phys.titech.ac.jp}}\\
\\
{\itshape Department of Physics, Tokyo Institute of Technology}, \\ {\itshape Tokyo 152-8551, Japan}}
\maketitle
\thispagestyle{empty}
\begin{abstract}
We study the di-baryon in the holographic QCD. The di-baryon is composed of six quarks and bound by the color interaction. In this paper, we adopt the Sakai-Sugimoto (SS) model as the holographic QCD to study the di-baryon. The SS model is formulated in a D4/D8/$\ol{\mathrm{D8}}$ system of the Type IIA string theory. It is expected that the di-baryon is described by the 2-instanton configuration of the flavor symmetry on D8-branes since the baryon number is identified as the instanton number. We will construct the 't Hooft 2-instanton solution explicitly and use it to discuss the stability of the di-baryon. An effective action of this model has not only the Yang-Mills (YM) action but also the Chern-Simons (CS) action coming from the CS action of probe D8-branes, and the CS action assigns $U(1)$ charges associated with the baryon number to each instanton. As a result, we can see that the di-baryon is unstable due to this $U(1)$ charge in the SS model. 
\end{abstract}
\end{titlepage}
\newpage
\section{Introduction}
The Sakai-Sugimoto model is one of the most successful holographic QCD models that describes low energy properties of the large $N_c$ QCD \cite{SS, SS2}. This model is formulated in a D4/D8/$\overline{\mathrm{D8}}$-brane system of the Type IIA string theory with an $S^1$ compactification of the fourth direction. The concrete configuration of this D-brane system is given in Table \ref{Dsystem}. \begin{table}[htbp]
\begin{center}
\begin{tabular}{|c|cccc|c|ccccc|}\hline
&0& 1 & 2 & 3 & 4 & 5 & 6 & 7 & 8 & 9 \\ \hline 
D4&$\circ $&$\circ $&$\circ $&$\circ $&$\circ $&&&&& \\ \hline 
D8-$\ol{\rm D8}$&$\circ $&$\circ $&$\circ $&$\circ $&&$\circ $& $\circ $&$\circ $&$\circ $&$\circ $ \\ \hline \hline
topology&\multicolumn{4}{|c|}{$M^4$}&$S^1$&\multicolumn{5}{|c|}{$\mathbb{R}^5$} \\ \hline
\end{tabular}
\caption{This is the D-brane configuration of the SS model in the ten dimensional space-time. The bottom line represents the topology of the D4 supergravity solution with the $S^1$ compactification of the forth direction.}
\label{Dsystem}
\end{center}
\end{table}
Note that the radius of this $S^1$ is $M_{\mathrm{KK}}^{-1}$, and the parameter $M_{\mathrm{KK}}$ that has a mass dimension is the typical energy scale of the model. The D4/D8/$\overline{\mathrm{D8}}$-brane system contains $N_c$ D4-branes corresponding to the color symmetry and $N_f$ D8-$\overline{\mathrm{D8}}$ pairs corresponding to the quark flavor symmetry. Since the model is based on the gauge/string duality that requires us to replace D4-branes with a D4 supergravity solution, the large $N_c$ limit is essential. The topology of the D4 supergravity solution with the $S^1$ compactification is $M^4\times S^1\times \mathbb{R}^5$ as given in Table \ref{Dsystem}. The DBI action and the CS action on $N_f$ D8-branes in the D4 supergravity solution give an effctive five dimensional YM-CS action for the $U(N_f)$ gauge field. This effective action represents mesons as Kaluza-Klein modes associated with the $S^1$ compactification. As a first application of this model to the realistic QCD, vector and scalar meson spectra are obtained from the effective action as eigenvalues of a Sturm-Liouville equation. Obtained results agree very well with experimental results, so that the SS model may be much powerful tool to understand low energy phenomena of the QCD.

The SS model can also describe baryons in two ways; the instanton description and the Skyrmion description. 

Actually the model includes the Skyrme model \cite{Skyrme,Skyrme2,Skyrme3} naturally. Since the Skyrme model is defined in the 4d space-time, we have to expand the 5d gauge field\footnote{Here we take the $A_z=0$ gauge.} $A_{\mu }(x,z)\in U(N_f)$ with respect to complete functions $\psi _n(z)$:
\begin{align}
A_{\mu }(x,z)=U^{-1}(x)\pd _{\mu }U(x)\psi _0(z)+\sum _{n=1}^{\infty }B_{\mu }^{(n)}(x)\psi _n(z),\label{expansion}
\end{align}
where $z$ is the direction associated with the $S^1$ compactification. The coefficient of the zero mode $U(x)\equiv \e ^{-\i \pi (x)/f_{\pi }}$ is the pion field and the coefficients of other modes $B_{\mu }^{(n)}(x)$ are the meson tower. If we write down an effective pion action using the expansion (\ref{expansion}), we get the Skyrme model with some vector meson contributions. This is called the brane-induced Skyrme model which has been studied in \cite{NSK}. In the phenomenological Skyrme model, baryons are described by soliton solutions like the hedgehog solution \cite{Skyrme3}, and therefore it is easy to expect that holographic baryons are also described by soliton solutions in the brane-induced Skyrme model. Since the effective pion action is constructed by using the expansion (\ref{expansion}), it is technically difficult to analyze the properties of baryons. 

On the other hand, in the instanton description, we use the 5d action directly. From the fact that baryons are described by the soliton solutions in the Skyrme model, it is quite natural to expect that holographic baryons are topological objects. In the SS model, a baryon is regarded as a D4-brane wrapping $S^4$ \cite{Witten}. This D4-brane is indeed interpreted as a topological gauge configuration on the probe D8-branes \cite{Douglas}, i.e., this gauge configuration is an instanton of the $U(N_f)$ gauge theory in the $(x^1,x^2,x^3,z)$-space. Furthermore, the instanton number is related to the baryon number. Under the curvature expansion for the D4 supergravity solution, the instanton configuration approximately satisfies the equations of motion obtained from the 5d effective YM-CS action. Later we will see this fact. The holographic baryons in the instanton description also has been studied in \cite{HSSY,HSS}.

In this paper, we would like to analyze di-baryons, not baryons, in the SS model. Di-baryons have the baryon number two. The YM action is proportional to the 't Hooft coupling $\lambda $, and the CS action does not have such a coupling. In the context of the gauge/string duality, since the curvature for the D4 supergravity solution is in inverse proportion to the 't Hooft coupling, the curvature expansion implies the $1/\lambda $ expansion. In this sense, the YM action is regarded as a leading term and the CS action is regarded as a sub-leading term. Actually di-baryons have been studied using the brane-induced Skyrme model in \cite{MNS,SM}. However, they considered only the leading contribution, that is, the YM action. In order to include the contribution of the CS action in the Skyrmion description, more complicated calculations are necessary due to the expansion (\ref{expansion}). Therefore we adopt the instanton description to analyze di-baryons with the YM-CS action in the SS model. Since the instanton number is interpreted as the baryon number, it is expected that a di-baryon is described as a 2-instanton configuration. The simplest case for the di-baryon analysis using instanton description is the two-flavor case, i.e., only $u$ and $d$ quarks. For instance as a two-flavor di-baryon, experimentally the $d^{\ast }(2380)$ di-baryon\footnote{This di-baryon is a $\Delta \Delta $ like object.} has been suggested \cite{Adlarson}. Hereafter, we will focus on such a two-flavor di-baryon and in particular, we will discuss the stability of the di-baryon. Unfortunately, we will see that the di-baryon is unstable due to the CS action.

The organization of this paper is as follows: In Sec. \ref{Preparation}, we will explain our notation and review roughly the SS model to analyze the di-baryon. In Sec. \ref{AD}, we will begin concrete analysis for the stability of the di-baryon. First we will treat only the YM action in the effective action. After this, we will analyze the YM-CS action and obtain the di-baryon potential as a function of moduli parameters of the 2-instanton. The analysis of this potential requires some numerical calculations for us. From these analyses, we will show that the CS action has a strong effect on the stability of the di-baryon. The conclusion is given in Sec. \ref{CD}. Throughout this paper, since we will use some knowledge of the 2-instanton many times, Appendix \ref{tHin} is devoted to the summary of the 2-instanton. 
\section{Preparation}\label{Preparation}
We will provide some ingredients to analyze di-baryon.
\subsection{Convention} 
First we should explain conventions in this paper. There are three types of indices:
\begin{itemize}
\item $\mu ,\nu ,\cdots =0,1,2,3$: four dimensional Minkowski space-time indicies,
\item $i,j,k,l=1,2,3$: three dimensional spatial indicies,
\item $m,n,p,q,r=1,2,3,z$: three dimensional spatial indicies and one dimensional internal space $z\in \mathbb{R}$ index\footnote{More precisely, $z$ is a combination of the forth direction and the fifth direction.}.
\end{itemize}
The Minkowski metric is $\eta ^{\mu \nu }={\rm diag}(-1,1,1,1)$. We will often encounter the combination of the coordinates $(x^i,z)$, so that it is useful to define $\xi ^m=(x^i,z)$. Then the norm $|\xi |$ means $\sqrt{x_i^2+z^2}$ and the summation over repeated indicies is understood. We will not distinguish the position of spatial indicies like 
\begin{align}
x_i&=x^i, & x_m&=x^m, 
\end{align}
because these have the Euclidean metric even if the space-time is curved: we will write the effect of the curved metric explicitly.

As mentioned in Introduction, we focus on the two-flavor di-baryon, so that the flavor symmetry on probe D8-branes is $U(2)$. We take a gauge field $A^{U(2)}$ of this $U(2)$ flavor symmetry as anti-Hermite, and therefore a gauge field of $U(1)$ part in this $U(2)$ is pure imaginary. 

We take a $M_{\rm KK}=1$ unit and it is easy to go back to the original natural unit by the dimensional analysis. 
\subsection{5d effective action of Sakai-Sugimoto model}  
In the $M_{\rm KK}=1$ unit, the 5d effective action of the SS model with the $U(2)=SU(2)\times U(1)$ flavor symmetry is given by
\begin{subequations}
\begin{align}
S&=S_{\rm YM}+S_{\rm CS},\\
S_{\rm YM}&=\kappa \int {\rm d}^4x{\rm d}z{\rm Tr}\left (\frac{1}{2}K^{-1/3}F_{\mu \nu }^2+KF_{\mu z}^2\right )\nonumber \\
&\qquad \quad +\frac{\kappa }{2}\int {\rm d}^4x{\rm d}z\left (\frac{1}{2}K^{-1/3}\widehat{F}_{\mu \nu }^2+K\widehat{F}_{\mu z}^2\right ),\\
S_{\rm CS}&=\frac{N_c}{24\pi ^2\i }\epsilon _{mnpq}\int {\rm d}^4x{\rm d}z\left [\frac{3}{8}\widehat{A}_0{\rm Tr}(F_{mn}F_{pq})-\frac{3}{2}\widehat{A}_m{\rm Tr}(\partial _0A_nF_{pq})\right .\nn \\
&\qquad \quad \left .+\frac{3}{4}\widehat{F}_{mn}{\rm Tr}(A_0F_{pq})+\frac{1}{16}\widehat{A}_0\widehat{F}_{mn}\widehat{F}_{pq}-\frac{1}{4}\widehat{A}_m\widehat{F}_{0n}\widehat{F}_{pq}\right ],\label{CSterm} \\
\kappa &=a\lambda N_c=\frac{\lambda N_c}{216\pi ^3},
\end{align}\label{action}
\end{subequations}
where $A$ and $\widehat{A}$ represent the $SU(2)$ and the $U(1)$ gauge field, respectively. The decomposition of $A^{U(2)}$ to these gauge fields is 
\begin{align}
A^{U(2)}=A+\frac{1}{2}\widehat{A}.
\end{align}
The additional factor $K\equiv 1+z^2$ is an effect of the curved background metric of the D4 supergravity solution. In this action, the first term of (\ref{CSterm})
\begin{align}
S_{\rm CS}\supset  \frac{N_c}{64\pi ^2{\rm i}}\epsilon _{mnpq}\int {\rm d}^4x{\rm d}z\widehat{A}_0{\rm Tr}(F_{mn}F_{pq})\label{U1charge}
\end{align}
has the role like the electric coupling. We can expect naively the di-baryon will be unstable due to this $U(1)$ coupling because each instanton has the same $U(1)$ charge associated with the baryon number. Later we will see that this $U(1)$ coupling indeed makes the di-baryon unstable.
\section{Analysis of Di-Baryon}\label{AD}
In this section, we will analyze the di-baryon concretely. First, we will make a potential of the di-baryon by using the YM action, not the YM-CS action, and try to find the minimum point of the potential. Next, we will also make the potential by using the YM-CS action. Then, it is more difficult to find the minimum point than the previous case, so that we need the numerical analysis for this process.
\subsection{Contribution from Yang-Mills action}
As the case of baryons \cite{HSSY}, the di-baryon should be realized as a solution that minimizes its potential. As mentioned in Introduction, it is expected that the di-baryon is described by the 2-instanton configuration of the flavor symmetry $U(2)$. Since later we will see that the 2-instanton configuration indeed satisfy the equations of motion approximately, we assume that the 2-instanton represents the di-baryon in the SS model. In general, the construction of the general 2-instanton configuration is technically difficult, even if we use the ADHM construction. The simplest 2-instanton is the 't Hooft 2-instanton which is fixed the $SU(2)$ orientation of the instanton, so that we adopt the 't Hooft instanton to describe the di-baryon. This 't Hooft 2-instanon configuration is given in (\ref{2inst}). Since we are interested in the static configuration, we neglect the time dependence of the 2-instanton configuration. Therefore, the potential of the di-baryon becomes as follows:
\begin{align}
V&=-\kappa \int {\rm d}^4\xi {\rm Tr}\left (\frac{1}{2}K^{-1/3}F_{ij}^2+KF_{iz}^2\right )\nonumber \\
&=\frac{\kappa }{4}\int {\rm d}^4\xi (K^{-1/3}+K)(-{\rm Tr}F_{mn}^2)\nonumber \\
&\geq \frac{\kappa }{2}\int {\rm d}^4\xi \sqrt{K^{-1/3}K}(-{\rm Tr}F_{mn}^2)\nonumber \\
&\geq \frac{\kappa }{2}\int {\rm d}^4\xi (-{\rm Tr}F_{mn}^2)=8\pi ^2\kappa N_B,\label{YMenergy}
\end{align}
with $K\geq K|_{z=0}=1$. Here we used formulas (\ref{usfom}) to factor out $-{\rm Tr}F_{mn}^2\geq 0$ in the first line. For our di-baryon case, the baryon number $N_B$ is equal to two. It is worth noting that this potential $V$ depends on ten moduli parameters of the 2-instanton configuration, i.e., the sizes $\rho _1,\rho _2$ and positions $a_1^m,a_2^m$ of each instanton. From the baryon result \cite{HSSY}, we can expect that the minimum point of the potential is given by the following configuration:
\begin{align}
\rho _1&=\rho _2=0, & a_1^z&=a_2^z=0.
\end{align}
To confirm this fact, let us estimate the integral in the second line of (\ref{YMenergy}) by using the Osborn's formula with above configuration. In this configuration, the Osborn's formula (\ref{Osborn}) becomes 
\begin{align}
{\rm Tr}F_{mn}^2=\partial _m^2\partial _n^2\log (\Xi _1^2\Xi _2^2)=-16\pi ^2\sum _{c=1,2}\delta ^{(3)}(x^i-a_c^i)\delta (z).
\end{align}
Then, the integral is easily estimated as
\begin{align}
V=-\frac{\kappa }{4}\int {\rm d}^4\xi (K^{-1/3}+K){\rm Tr}F_{mn}^2=16\pi ^2\kappa .
\end{align}
This result indeed coincides with the last line of (\ref{YMenergy}) for $N_B=2$. It is quite easy to generalize this result to an arbitrary 't Hooft $k$-instanton, where $k=N_B$ and the minimum point is given by $\rho _c=0$ and $a_c^z=0$. Then, the minimum potential energy is reproduced by this configuration immediately. In summary of this section, the curved background metric affects only the $a_1^z,a_2^z$ dependence of the potential, and instantons concentrate on the origin in the $z$ direction. Therefore, the conclusion of this section is that the stability of the di-baryon is marginal if we consider only the YM action. In the next section we will refine this result by including the CS action.
\subsection{Contribution including Chern-Simons action}
Now we have seen that the stability of the di-baryon is marginal, so that the di-baryon is not bound state if we consider only the YM action. Since there is also the CS action in SS model, it is important to investigate the role of this term. As we mentioned before, there is the electric coupling term in the CS action, so that naively we have expected that the di-baryon is unstable due to the $U(1)$ coupling. We will estimate this effect quantitatively below.

Ingredients such as equations of motion necessary for our di-baryon analysis have already been given in \cite{HSSY}, thus we will give a short explanation for the derivation of equations of motion. In the case of the baryon analysis, the size at the minimum point configuration is of order $\lambda ^{-1/2}$ in the context of the $1/\lambda $ expansion. Thus we expect that sizes of both instantons in the 2-instanton are also of order $\lambda ^{-1/2}$. Then, it is convenient to extract the factor $\lambda ^{-1/2}$ from $x^m$ and $\lambda ^{1/2}$ from $A_m,\widehat{A}_m$:
\begin{align}
\begin{pmatrix}x^0&x^m\\ A_0&A_m\\ \widehat{A}_0&\widehat{A}_m\end{pmatrix}\to \begin{pmatrix}x^0&\lambda ^{-1/2}x^m\\ A_0&\lambda ^{1/2}A_m\\ \widehat{A}_0&\lambda ^{1/2}\widehat{A}_m\end{pmatrix}.\label{extraction}
\end{align}
This extraction enables us to carry out $1/\lambda $ expansion in much simpler way. As a result, the action up to $O(\lambda ^{-1})$ is obtained as
\begin{align}
S&=aN_c\int \d ^4x\d z\bigg [{\rm Tr}\skakko{\frac{\lambda }{2}F_{mn}^2-\frac{z^2}{6}F_{ij}^2+z^2F_{iz}^2-F_{0m}^2}\nonumber \\
&\quad +\frac{1}{2}\skakko{\frac{\lambda }{2}\widehat{F}_{mn}^2-\frac{z^2}{6}\widehat{F}_{ij}^2+z^2\widehat{F}_{iz}^2-\widehat{F}_{0m}^2}\bigg ]\nn \\
&\quad +\frac{N_c}{24\pi ^2\i }\epsilon _{mnpq}\int \d ^4x\d z\bigg [\frac{3}{8}\widehat{A}_0{\rm Tr}(F_{mn}F_{pq})-\frac{3}{2}\widehat{A}_m{\rm Tr}(\pd _0A_nF_{pq})\nn \\
&\qquad \qquad \qquad +\frac{3}{4}\widehat{F}_{mn}{\rm Tr}(A_0F_{pq})+\frac{1}{16}\widehat{A}_0\widehat{F}_{mn}\widehat{F}_{pq}-\frac{1}{4}\widehat{A}_m\widehat{F}_{0n}\widehat{F}_{pq}\bigg ]+O(\lambda ^{-1}).\label{action2}
\end{align}
Here note that the CS action is invariant under the extraction (\ref{extraction}). The leading part of equations of motion obeyed from this action are \cite{HSSY}
\begin{subequations}
\begin{align}
D_mF_{mn}=0,\label{inst}\\
\partial _m\widehat{F}_{mn}=0,\label{pure}\\
D_mF_{0m}-\frac{1}{64\pi ^2{\rm i}a}\epsilon _{mnpq}\widehat{F}_{mn}F_{pq}=0,\label{noneed}\\
\partial _m\widehat{F}_{0m}-\frac{1}{64\pi ^2{\rm i}a}\epsilon _{mnpq}\left [{\rm Tr}(F_{mn}F_{pq})+\frac{1}{2}\widehat{F}_{mn}\widehat{F}_{pq}\right ]=0.\label{eqos}
\end{align}\label{eom}
\end{subequations}
Let us find solutions corresponding to the di-baryon. The procedure is as follows:
\begin{enumerate}
\item Eq. (\ref{inst})

This equation shows that the instanton configuration is approximately a solution in $1/\lambda $ expansion. Therefore, this fact justifies to use the 2-instanton configuration as the di-baryon shown in (\ref{2inst}). 
\item Eq. (\ref{pure})

Since we would like to obtain the finite energy solution, $F_{mn}=0$. This equation merely gives a pure gauge configuration, $\widehat{A}_m=0$. 
\item Eq. (\ref{noneed})

Combining the result of this pure gauge configuration with the time independence of instanton configuration, we can reduce Eq. (\ref{noneed}) to 
\begin{align}
D_n^2A_m=0.\label{noneed2}
\end{align}
But there is no need to solve this equation. 
\item Eq. (\ref{eqos})

In a similar way that reduces Eq. (\ref{noneed}), Eq. (\ref{eqos}) is reduced to 
\begin{align}
\partial _m^2\widehat{A}_0=-\frac{1}{32\pi ^2{\rm i}a}{\rm Tr}(F_{mn}^2).\label{reos}
\end{align}
It is easy to solve this equation by comparing it with Osborn's formula (\ref{Osborn}). The solution that vanishes at infinity is given by
\begin{align}
\widehat{A}_0=-\frac{1}{32\pi ^2{\rm i}a}\partial _m^2\log \Delta .
\end{align} 
\end{enumerate}

Now we have obtained some solutions of Eqs. (\ref{eom}) to calculate the potential of the di-baryon. To get this potential, let us substitute these solutions to the action (\ref{action2}). In this process, as we mentioned before, Eq. (\ref{noneed2}) can be used to drop the term ${\rm Tr}(F_{0m}^2)$ in the action. The self dual condition for the field strength is also useful in this calculation. After integrating by part and comparing the action with the on-shell action $S=-\int {\rm d}x^0\mathcal{M}$, we find the potential of the di-baryon
\begin{align}
\mathcal{M}=16\pi ^2\kappa -\frac{aN_c}{6}\int {\rm d}^4\xi z^2{\rm Tr}(F_{mn}^2)-\frac{N_c}{64\pi ^2{\rm i}}\int {\rm d}^4\xi \widehat{A}_0{\rm Tr}(F_{mn}^2).\label{mass}
\end{align} 
As we have already known, this potential depends on ten moduli parameters of the 2-instanton solution. We would like to find the stability point of this potential, i.e., minimum point. Before starting the concrete analysis, we should refer to a relation between the 1-instanton and the 2-instanton. In fact, if we take the limit\footnote{In the component style, $\boldsymbol{a}_2$ means $a_2^i$.} $|\bv{a}_2|\to \infty $ for the trace square of the field strength, the 1-instanton result is reproduced immediately\footnote{For simplicity, we assume $\bv{a}_1=0$ for this calculation. This assumption is justified by the translation symmetry of the system.}:
\begin{align}
\lim _{|\boldsymbol{a}_2|\to \infty }{\rm Tr}(F_{mn}^2)=-\frac{96\rho _1^4}{[r^2+(z-a_1^z)^2+\rho _1^2]^4},\label{2lim}
\end{align}
where we use the explicit form of field strength of the 2-instanton (\ref{trsq}). Here $r=\sqrt{x_i^2}$ is the spatial distance. This form precisely coincides with 1-instanton result, so that if we substitute (\ref{2lim}) to the potential of the di-baryon (\ref{mass}), the baryon result shown in \cite{HSSY} is immediately reproduced after the $\xi $-integral.

Let us find the stability point in the 2-instanton moduli space by adjusting its moduli parameters. Since the potential (\ref{mass}) is given by the integral form, we need numerical calculations to realize our desire. Now we are interested in the part of the binding energy $M$:
\begin{subequations} 
\begin{align}
\mathcal{M}&=16\pi ^2\kappa +N_cM,\\
M&=-\frac{a}{6}\int {\rm d}^4\xi z^2{\rm Tr}(F_{mn}^2)-\frac{1}{64\pi ^2{\rm i}}\int {\rm d}^4\xi \widehat{A}_0{\rm Tr}(F_{mn}^2).
\end{align}
\end{subequations}
Thus, numerical results will be shown only the part $M$ of $\mathcal{M}$. Despite the ten moduli parameters dependence, these parameters can be reduced to only three parameters without loss of generality due to an exchanging symmetry between instantons. More strictly speaking, the potential only depends on instanton sizes, a distance to each other in spatial direction and each positions along the $z$ axis. Therefore, when analyzing the potential, we only consider following three parameters:
\begin{itemize}
\item the size of both instantons $\rho _1=\rho _2\equiv \rho $,
\item the distance to each other in spatial direction $|\boldsymbol{a}_1-\boldsymbol{a}_2|\equiv |\boldsymbol{a}_{12}|$,
\item the position along the $z$ axis $a_1^z=-a_2^z\equiv a_z$.
\end{itemize}
\begin{figure}[htbp]
\begin{minipage}{0.5\hsize}
\begin{center}
\includegraphics[width=110mm, bb=0 0 580 300]{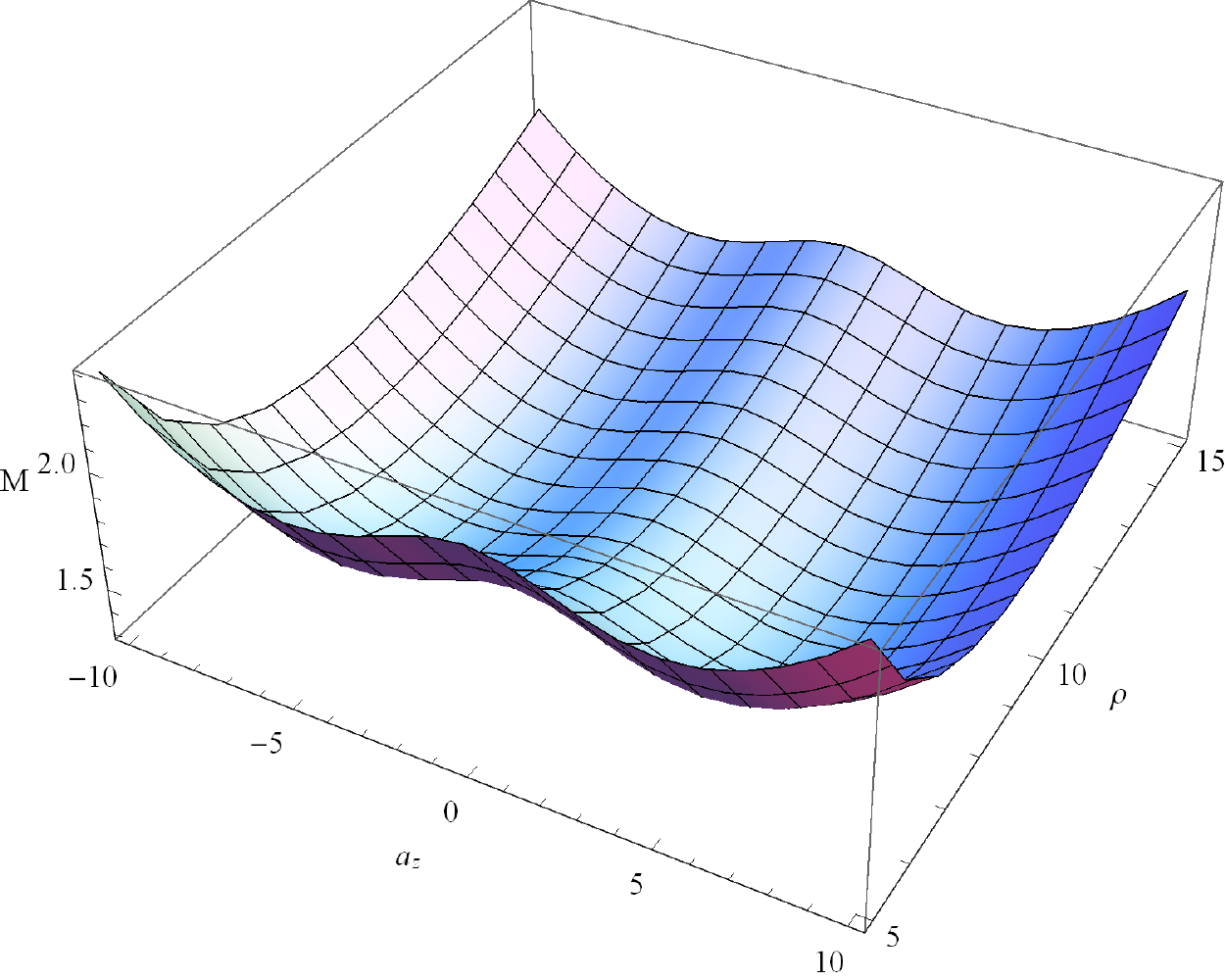}
\hspace{0.7cm} $|\bv{a}_{12}|=10$
\end{center}
\end{minipage}
\begin{minipage}{0.5\hsize}
\begin{center}
\includegraphics[width=110mm, bb=0 0 580 300]{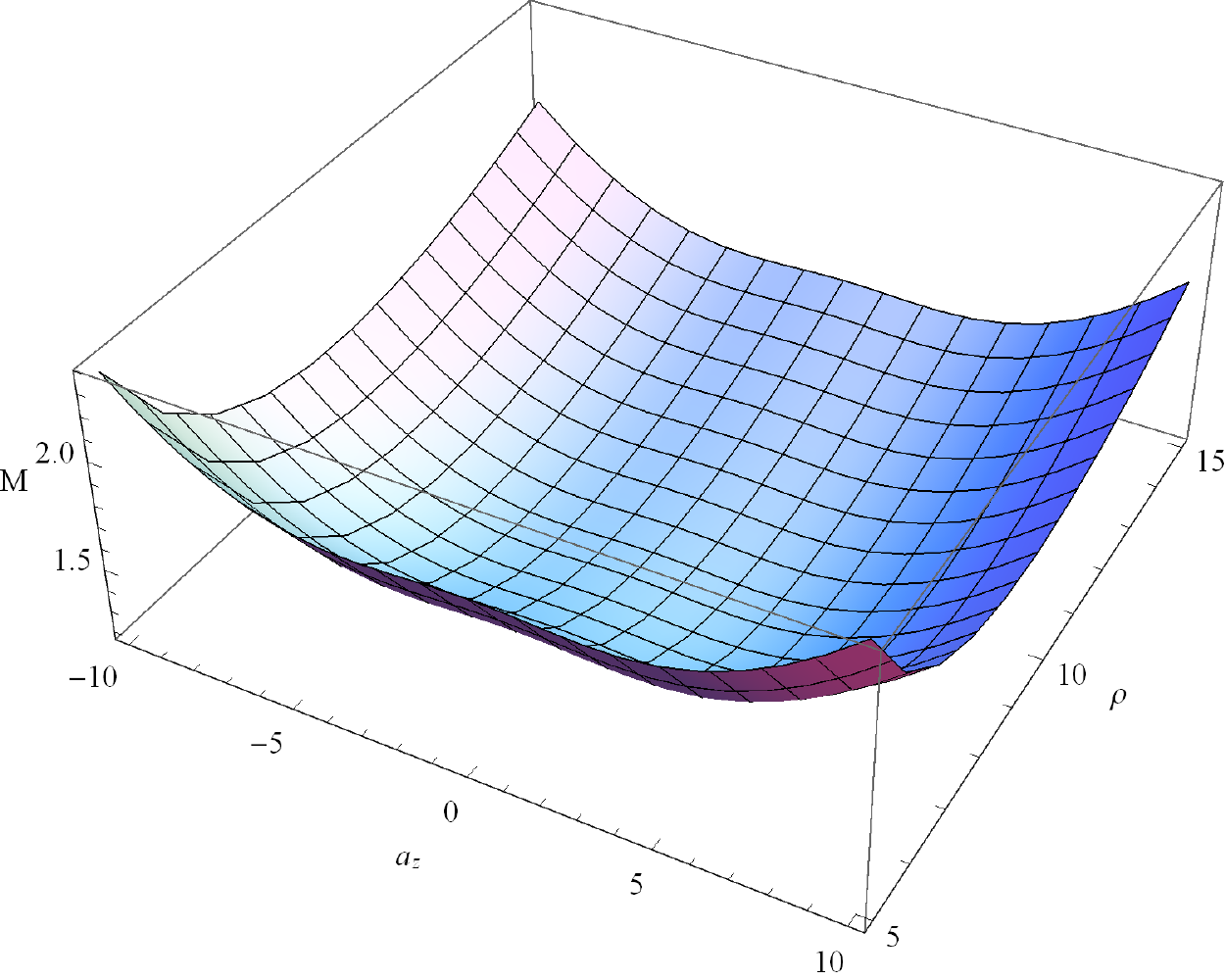}
\hspace{0.7cm} $|\bv{a}_{12}|=12.5$
\end{center}
\end{minipage}
\begin{minipage}{0.5\hsize}
\begin{center}
\includegraphics[width=110mm, bb=0 0 580 300]{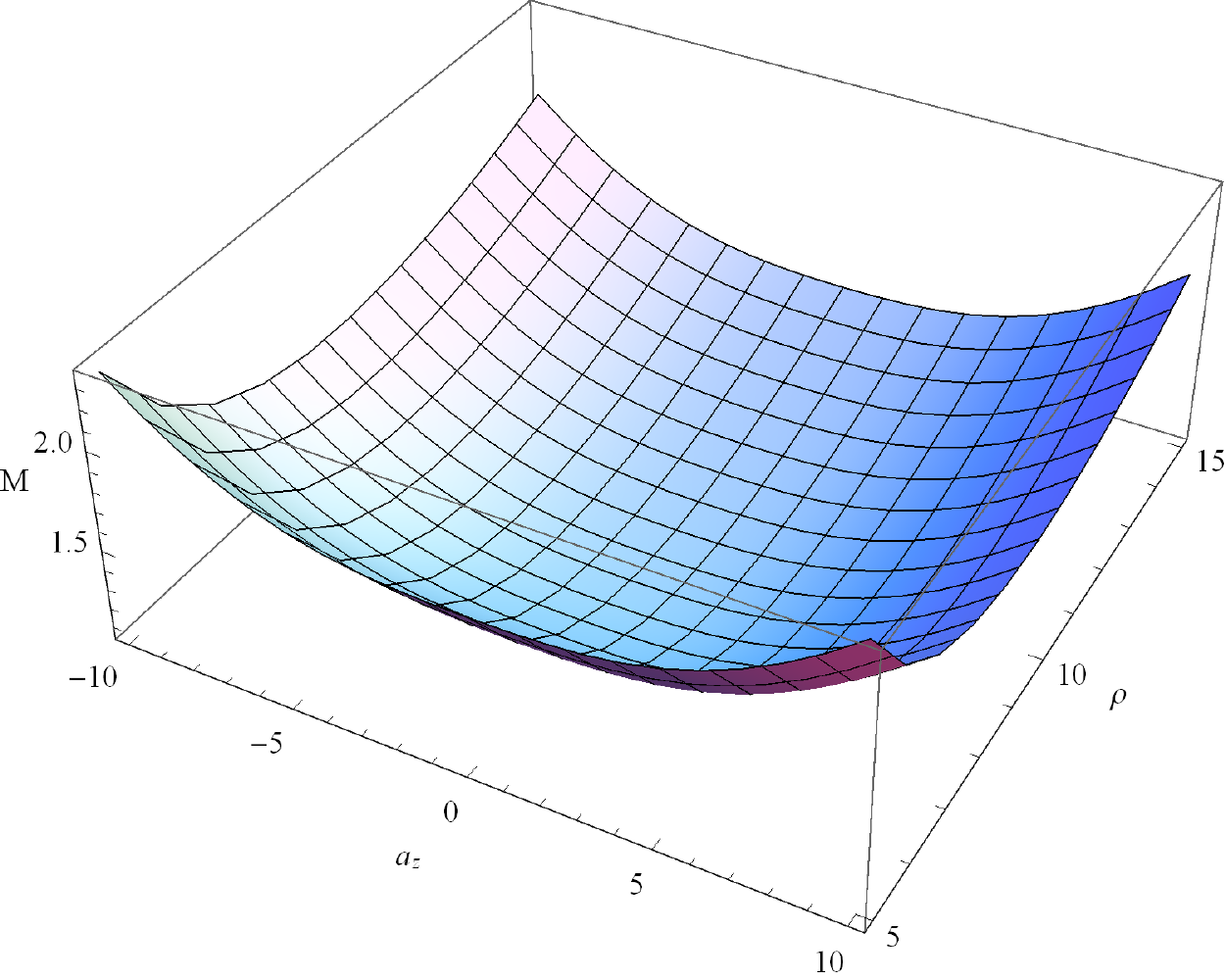}
\hspace{0.7cm} $|\bv{a}_{12}|=15$
\end{center}
\end{minipage}
\begin{minipage}{0.5\hsize}
\begin{center}
\includegraphics[width=110mm, bb=0 0 580 300]{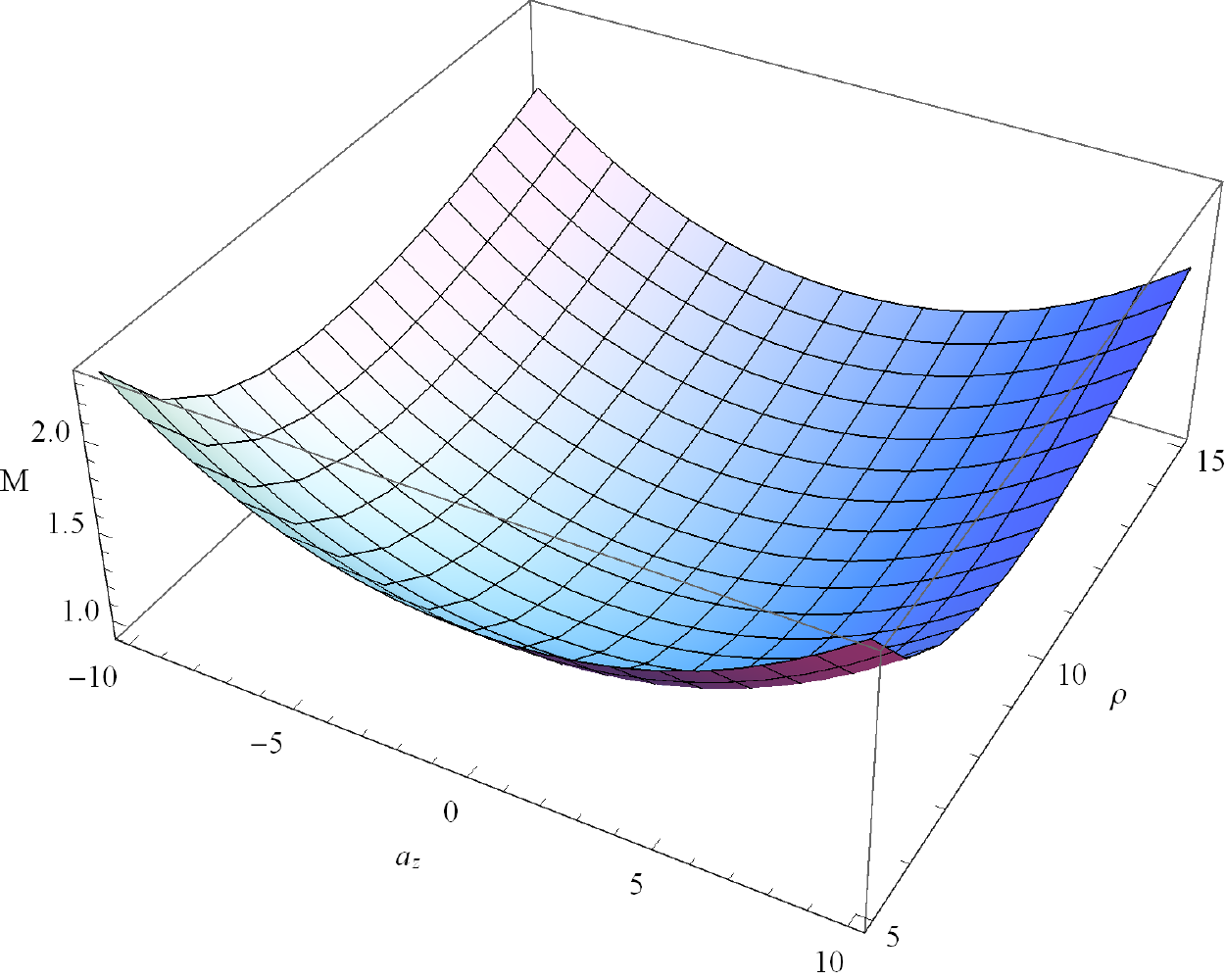}
\hspace{0.7cm} $|\bv{a}_{12}|=20$
\end{center}
\end{minipage}
\begin{minipage}{0.5\hsize}
\begin{center}
\includegraphics[width=110mm, bb=0 0 580 300]{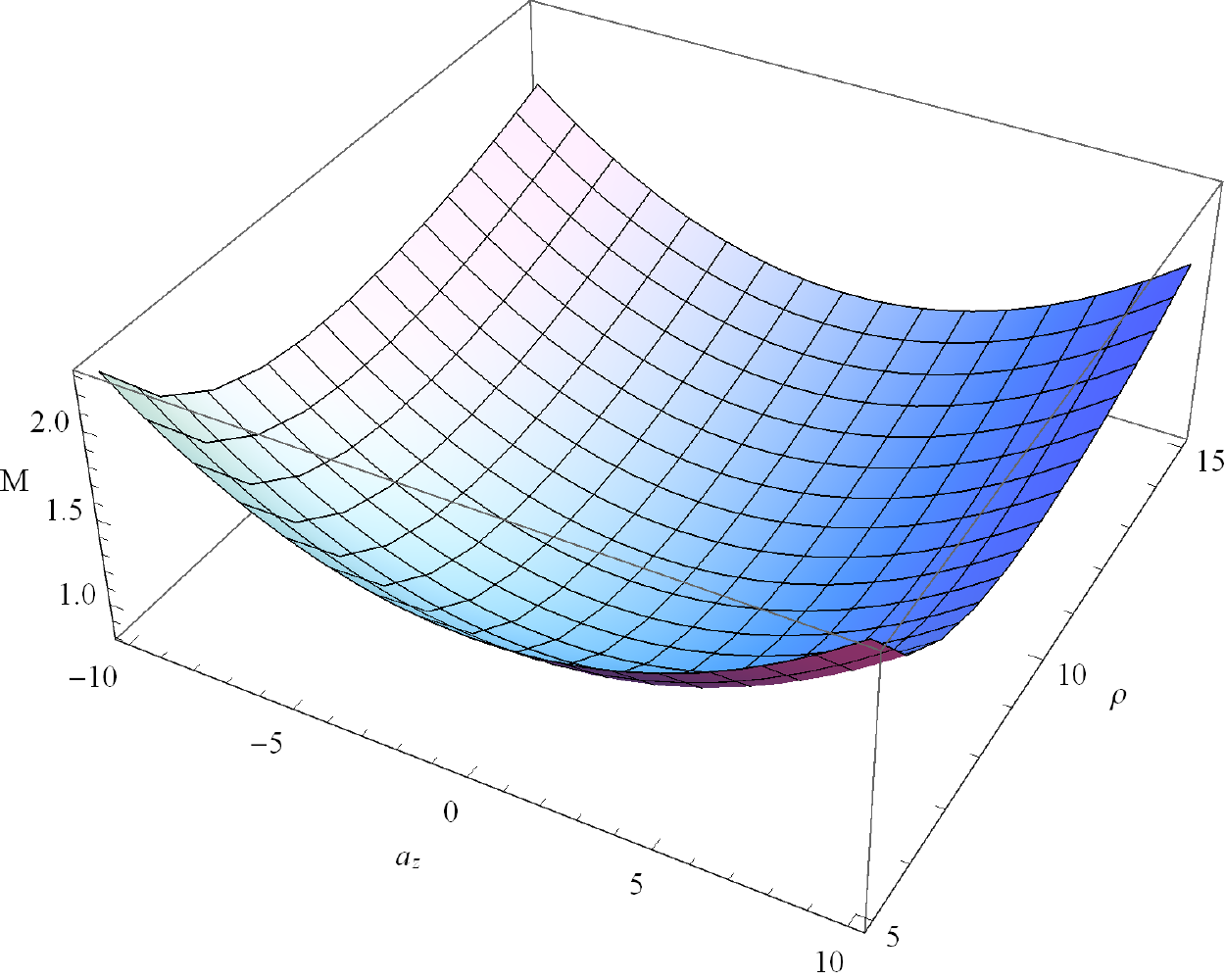}
\hspace{0.7cm} $|\bv{a}_{12}|=30$
\end{center}
\end{minipage}
\begin{minipage}{0.5\hsize}
\begin{center}
\includegraphics[width=110mm, bb=0 0 580 300]{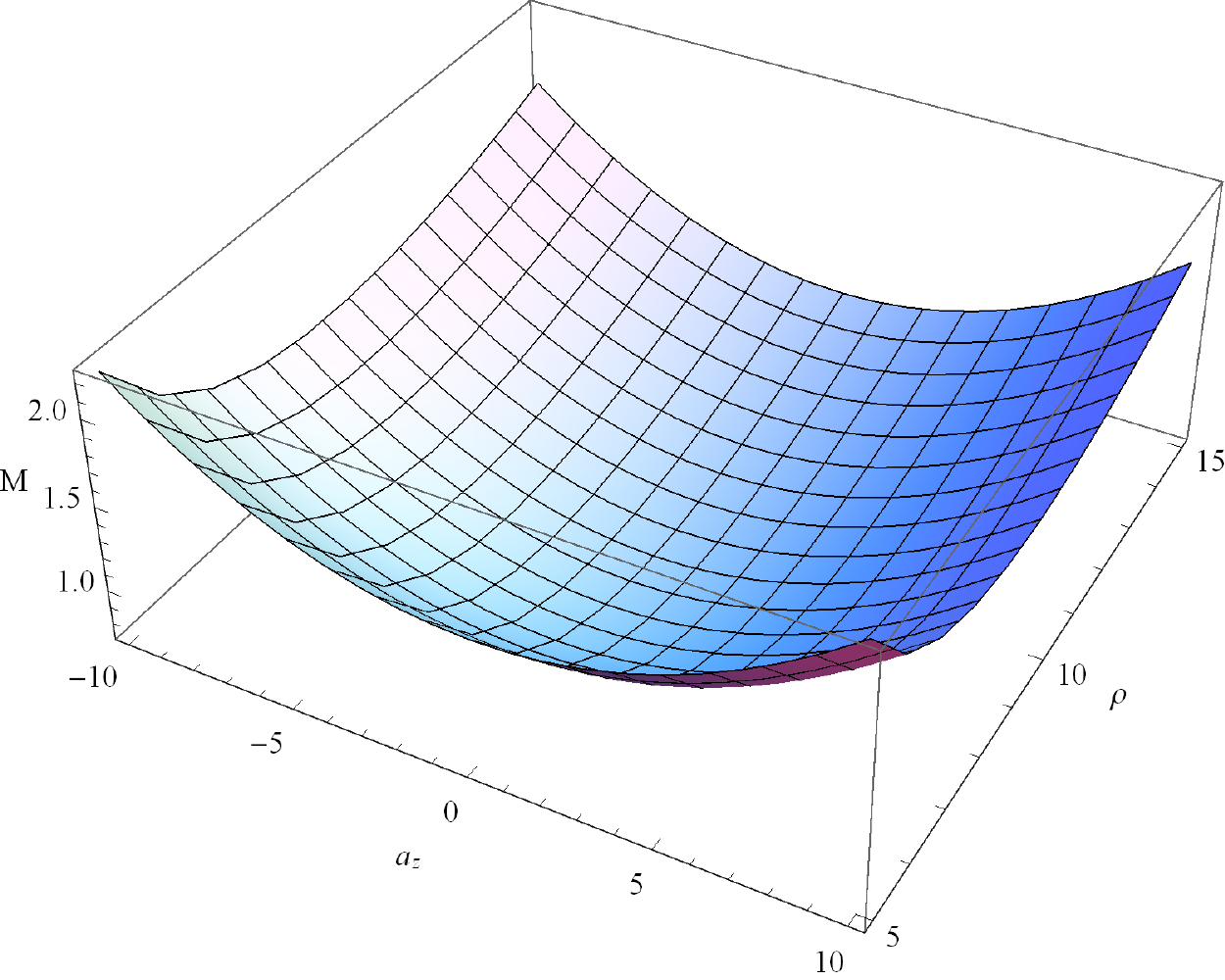}
\hspace{0.7cm} $|\bv{a}_{12}|\to \infty $
\end{center}
\end{minipage}
\caption{Plots of $M$ in the parameter space $(\rho ,a_z)$ at various $|\boldsymbol{a}_{12}|$. The analyzing range is determined by the 1-instanton and the baryon result \cite{HSSY}. In the baryon case, the minimum value is $(\rho ,a_z)\sim (10,0)$, thus we can focus on the range near this point for the di-baryon analysis.}\label{a10}
\end{figure}
\begin{figure}[htbp]
\begin{center}
\includegraphics[width=220mm, bb=0 0 580 300]{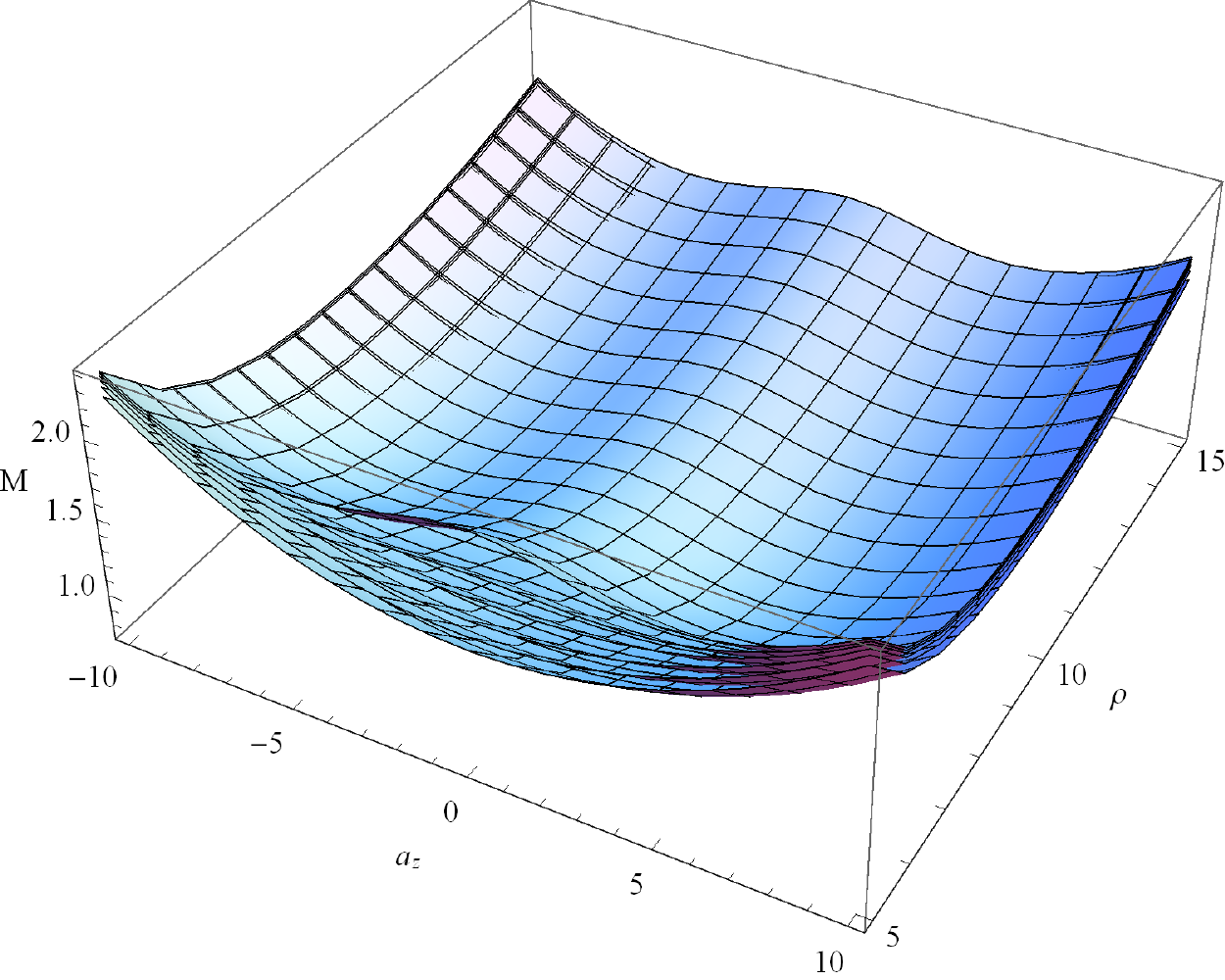}
\end{center}
\caption{A collection of all graphs in Fig. \ref{a10}. The top graph is the $|\bv{a}_{12}|=10$ case, the second is the $|\bv{a}_{12}|=12.5$ case and so on. The last graph is the $|\bv{a}_{12}|\to \infty $ case. This ordering shows that the value of potential is monotonically decreasing along the $|\boldsymbol{a}_{12}|$ axis and graphs for finite $|\boldsymbol{a}_{12}|$ asymptotically approach to the $|\boldsymbol{a}_{12}|\to \infty $ case.}\label{all}
\end{figure}

Now we are ready for analyzing the potential numerically. Results are shown in Fig. \ref{a10} and Fig. \ref{all}. These figures show that the potential as a function of $\rho ,a_z$ at each value of $|\boldsymbol{a}_{12}|$. In Fig. \ref{a10}, the vertical line is the potential and we can see two stable points on the $(\rho ,a_z)$-plane. The investigating range of $(\rho ,a_z)$ is determined by the 1-instanton result \cite{HSSY}: since the minimum point is $(\rho ,a_z)=(9\pi (15\pi ^2/2)^{-1/4},0)\sim (10,0)$ in baryon case, we only see the range near this point. Since we would like to discuss the stability of the di-baryon, the position of the minimum point is not important. From Fig. \ref{a10}, we can also read the feature that two stable points on the $a_z$ axis are shifting to origin as $|\bv{a}_{12}|$ increases. This shift has a physical meaning: since instantons can not approach to each other due to the $U(1)$ interaction in the CS action, stable points along the $a_z$ axis can not exist near the origin when the value of $|\boldsymbol{a}_{12}|$ is small. The existence of the stable point along the $\rho $ axis at the finite value is due to the same reason as electrodynamics: the energy of an infinitesimal size charged particle is diverge. The $|\bv{a}_{12}|\to \infty $ case in Fig. \ref{a10} is quite important. In this limit, the $U(1)$ interaction becomes irrelevant. Then the contribution to the potential is only two times of the 1-instanton given in \cite{HSSY}. Finally, Fig. \ref{all} shows all figures in Fig. \ref{a10}. What we can understand from Fig. \ref{all} is the value of the potential along the $|\boldsymbol{a}_{12}|$ axis monotonically decreases and approaches to the $|\boldsymbol{a}_{12}|\to \infty $ case in Fig. \ref{a10}. This implies that there is no stable point along the $|\boldsymbol{a}_{12}|$ axis, so that we conclude that the di-baryon is unstable object in the SS model if we use the 't Hooft ansatz for instantons. The instability of the di-baryon is also due to the $U(1)$ interaction. As we mentioned before, our naive expectation about the stability of the di-baryon is therefore correct.
\section{Conclusion}\label{CD}
Throughout this paper, we have studied the stability of the di-baryon and showed that the di-baryon is unstable in the SS model in the previous section. Now we will give a short summary. First we have considered the contribution to the potential from the YM action only. Then we found that the stability of the di-baryon is marginal. The minimum value of the potential already have been derived in \cite{HSSY}. Second, we saw the contribution from the YM-CS action. In this case, we needed numerical calculations to investigate whether the di-baryon is stable or unstable. Before performing numerical computations, we took the limit $|\boldsymbol{a}_2|\to \infty $ and reproduced the 1-instanton result \cite{HSSY}. The numerical result showed that the $|\boldsymbol{a}_{12}|$ dependence of the potential is monotonically decreasing, so that the marginal stability of the di-baryon is broken by the CS action. By combining results with the YM action case, we can conclude that this instability along the $|\boldsymbol{a}_{12}|$ axis is due to the $U(1)$ interaction in the CS action. 
\section*{Acknowledgments}
I would like to thank Yosuke Imamura, Makoto Oka, Takafumi Okubo and Shigeki Sugimoto for variable discussions. I also acknowledge the financial support from International Research Center for Nanoscience and Quantum Physics, Tokyo Institute of Technology. 
\appendix
\section{'t Hooft 2-Instanton Gauge Configuration for $SU(2)$}\label{tHin}
In this appendix, we will show the 't Hooft 2-instanton configuration for $SU(2)$. For more details, see \cite{NTVM,WY,Tian}. The 't Hooft $k$-instanton gauge configuration for $SU(2)$ is given by the 't Hooft ansatz with coordinates $\xi ^m \in \mathbb{R}^4$:
\begin{subequations}
\begin{align}
A_m&=\frac{\rm i}{2}\eta _{mn}^{(-)}\partial _{n}\log \Phi =-\frac{\rm i}{\Phi }\eta _{mn}^{(-)}\sum _{c=1}^k\frac{\rho _c^2(\xi -a_c)^n}{|\xi -a_c|^4},\\
\Phi &=1+\sum _{c=1}^k\frac{\rho _c^2}{|\xi -a_c|^2},
\end{align}\label{2inst}
\end{subequations}
where $\eta _{mn}^{(-)}$ is the 't Hooft eta symbol defined as
\begin{align}
\eta _{mn}^{(\pm )}=\eta _{mn}^{i(\pm )}\sigma ^i=(\epsilon _{imnz}\pm \delta _{im}\delta _{nz}\mp \delta _{in}\delta _{mz})\sigma ^i,
\end{align}
and $\sigma ^i$ is the standard Pauli matrices. Note that this 't Hooft $k$-instanton configuration has $5k$ moduli parameters: each instanton size $\rho _c$ and each instanton position $a^m_c$. We will use the $k=2$ case to represent the di-baryon in the SS model, and in this case we can calculate the field strength explicitly:
\begin{subequations}
\begin{align}
F_{mn}&=-\frac{2{\rm i}}{\Delta ^2}\Xi _1^2\Xi _2^2\eta _{mn}^{i(+)}(\delta _{ij}Y-2Y_{ij}+2\epsilon _{ijpz}Y_{pz}-2\delta _{ij}Y_{zz})\sigma ^j,\\
Y^{mn}&=(\rho _2^2+\Xi _2^2)\frac{\rho _1^2\Xi _1^m\Xi _1^n}{\Xi _1^4}+(\rho _1^2+\Xi _1^2)\frac{\rho _2^2\Xi _2^m\Xi _2^n}{\Xi _2^4}-\frac{\rho _1^2\rho _2^2}{\Xi _1^2\Xi _2^2}(\Xi _1^m\Xi _2^n+\Xi _1^n\Xi _2^m),\\
\Delta &=\rho _1^2\Xi _2^2+\rho _2^2\Xi _1^2+\Xi _1^2\Xi _2^2=\Xi _1^2\Xi _2^2\Phi ,
\end{align}
\end{subequations}
where $\Xi _c^m=\xi ^m-a_c^m$ and $Y$ is the trace of $Y^{mn}$, i.e., $Y=Y^{mm}$. Using these results, one can compute the trace square of the field strength as follows:
\begin{align}
{\rm Tr}(F_{mn}^2)=-96\left (\frac{\rho _1^2\Xi _2^4+\rho _2^2\Xi _1^4+a_{12}^2\rho _1^2\rho _2^2}{\Delta ^2}\right )^2+256\rho _1^2\rho _2^2\frac{\Xi _1^2\Xi _2^2-(\Xi _1\cdot \Xi _2)^2}{\Delta ^3},\label{trsq}
\end{align}
where $a_{12}^m=a_1^m-a_2^m$ is the difference between instanotn positions. In this computation, we used some useful formulas
\begin{align}
\eta _{mn}^{i(+)}\eta _{mn}^{j(+)}&=4\delta ^{ij}, & \eta _{kl}^{i(+)}\eta _{kl}^{j(+)}&=2\delta ^{ij}, & \eta _{kz}^{i(+)}\eta _{kz}^{j(+)}=\delta ^{ij}.
\end{align}
Furthermore, these useful formulas imply that
\begin{align}
{\rm Tr}(F_{ij}^2)&=\frac{1}{2}{\rm Tr}(F_{mn}^2), & {\rm Tr}(F_{iz}^2)&=\frac{1}{4}{\rm Tr}(F_{mn}^2).\label{usfom}
\end{align}

Finally, we should mention the Osborn's formula
\begin{align}
{\rm Tr}(F_{mn}^2)=\partial _m^2\partial _n^2\log \Delta ,\label{Osborn}
\end{align}
which is followed from the ADHM construction. Using this formula, we can immediately reproduce the 2-instanton number:
\begin{align}
-\frac{1}{32\pi ^2}\epsilon _{mnpq}\int {\rm d}^4\xi {\rm Tr}(F_{mn}F_{pq})=2.\label{instn}
\end{align}
This instanton number is also reproduced by the numerical integral of the left hand side of (\ref{instn}).

\end{document}